\documentclass[runningheads]{llncs}
\usepackage[T1]{fontenc}
\usepackage{graphicx}
\usepackage{tabularx, multirow, float}
\usepackage{placeins}
\usepackage{csquotes}
\usepackage{eurosym}

\usepackage{url}
\usepackage{ifthen}
\usepackage{url}
\usepackage{breakurl}
\usepackage[super]{nth}
\usepackage[breaklinks]{hyperref}
\usepackage{advdate}

\usepackage{siunitx}
\usepackage{soul}

\usepackage{graphicx}

\usepackage[labelformat=simple]{subcaption}

\usepackage{multirow}

\usepackage{cleveref}
\usepackage{booktabs}
\usepackage{multirow}

\usepackage{svg}

\usepackage{xcolor}
\usepackage{mdframed}

\usepackage{comment}

\definecolor{siemens}{HTML}{ec6602}

\usepackage[misc]{ifsym}

\DeclareSIUnit\voxel{voxel}

\usepackage{hyperref}



\begin{document}
\title{Towards Integrating Epistemic Uncertainty Estimation into the Radiotherapy Workflow}

\titlerunning{Uncertainty Estimation in Radiotherapy}


%
%
\author{Marvin Tom Teichmann\inst{1}\textsuperscript{(\Letter)} \and
Manasi Datar\inst{1} \and
Lisa Kratzke\inst{2} \and Fernando Vega\inst{2} \and Florin C. Ghesu\inst{1} \\\vspace{0.5em} \email{
marvin.teichmann@siemens-healthineers.com}}
\authorrunning{M. T. Teichmann et al.}

 \institute{
 Digital Technology and Innovation, Siemens Healthineers, Erlangen, Germany
\and Cancer Therapy Imaging, Varian, Siemens Healthineers, Forchheim, Germany
 }


%
%
\maketitle              
\begin{abstract}
The precision of contouring target structures and organs-at-risk (OAR) in radiotherapy planning is crucial for ensuring treatment efficacy and patient safety. Recent advancements in deep learning (DL) have significantly improved OAR contouring performance, yet the reliability of these models, especially in the presence of out-of-distribution (OOD) scenarios, remains a concern in clinical settings. This application study explores the integration of epistemic uncertainty estimation within the OAR contouring workflow to enable OOD detection in clinically relevant scenarios, using specifically compiled data. Furthermore, we introduce an advanced statistical method for OOD detection to enhance the methodological framework of uncertainty estimation. Our empirical evaluation demonstrates that epistemic uncertainty estimation is effective in identifying instances where model predictions are unreliable and may require an expert review. Notably, our approach achieves an AUC-ROC of \num{0.95} for OOD detection,  with a specificity of \num{0.95} and a sensitivity of \num{0.92} for implant cases, underscoring its efficacy. This study addresses significant gaps in the current research landscape, such as the lack of ground truth for uncertainty estimation and limited empirical evaluations. Additionally, it provides a clinically relevant application of epistemic uncertainty estimation in an FDA-approved and widely used clinical solution for OAR segmentation from Varian, a Siemens Healthineers company, highlighting its practical benefits.
\end{abstract}

The advent of precision medicine in radiotherapy has underscored the importance of accurate contouring of target structures and organs at risk (OAR) on computed tomography (CT) simulation scans. This critical step ensures that the dose objectives set by oncologists are met, directly influencing treatment outcomes. Recent advancements in DL have significantly enhanced the accuracy of contouring tasks, offering promising solutions to the challenges of manual delineation, which is time consuming, costly and suffers from inter-observer variability \cite{joskowicz2019inter}. Despite these advancements, the reliability of DL models in clinical settings is often questioned, particularly when faced with OOD scenarios that were not represented in the training data. Such scenarios can lead to inaccuracies in contouring, posing significant risks to patient safety. In this application study we employ uncertainty estimation to address the challenges posed by OOD data. Specifically, we aim to evaluate the utility of uncertainty estimation for OOD detection, a critical aspect where traditional model confidences may fall short. Identification of OOD instances early in the radiotherapy workflow can help mitigate potential errors in the ensuing steps. 

A recent review \cite{zou2023review} examines the current research landscape in uncertainty estimation for medical imaging applications, highlighting its pivotal role in enhancing the reliability and trustworthiness of DL models in healthcare. Although it is an actively pursued area of research, the review identifies two major obstacles hindering progress and adaption: the
lack of ground truth data and tasks for uncertainty estimation, and the limited empirical evaluations of uncertainty estimation methods, especially for real clinical applications. This study aims to address these aforementioned challenges within the realm of epistemic uncertainty estimation. To achieve this, we identify clinically relevant OOD scenarios for the OAR contouring task in radiotherapy. We then gather corresponding datasets and conduct empirical evaluations to assess the effectiveness of epistemic uncertainty estimation in those scenarios.

The contributions of our application study of uncertainty estimation for radiotherapy contouring are threefold: Firstly, we demonstrate the utility of uncertainty estimation for OOD detection by identifying clinically relevant use-cases and showing its effectiveness in mitigating associated risks. Secondly, we introduce a new OOD detection method using advanced statistical analysis for threshold estimation. Lastly, we aim to bridge the gap between research and clinical practice by showing how uncertainty estimation can provide quantifiable improvements for review of DL-contouring results in an FDA-approved and widely used pipeline for OAR segmentation by Varian, a Siemens Healthineers company. Our primary goal is to highlight the practical benefits of uncertainty estimation, and we aim to inspire further comparative evaluations on public datasets and advance the field of uncertainty quantification in medical imaging.

\section{Our Datasets for Uncertainty Model Evaluation}

\begin{figure*}[t]
\centering
\begin{subfigure}[t]{0.33\textwidth}
\centering
\includegraphics[width=\textwidth]{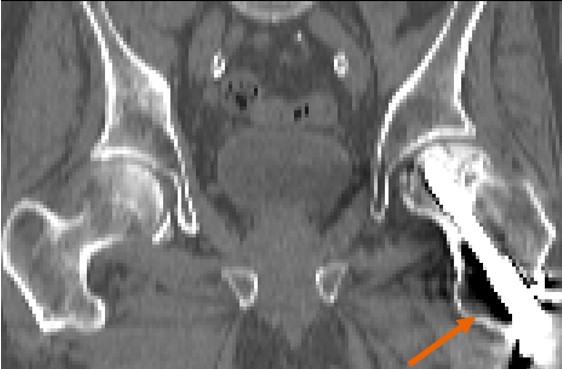}
\caption{}
\label{fig:femur}
\end{subfigure}%
~
\begin{subfigure}[t]{0.33\textwidth}
\centering
\includegraphics[width=\textwidth]{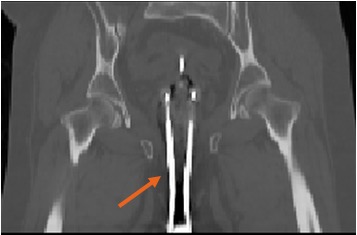}
\caption{}
\label{fig:brachy}
\end{subfigure}%
~
\begin{subfigure}[t]{0.33\textwidth}
\centering
\includegraphics[width=\textwidth]{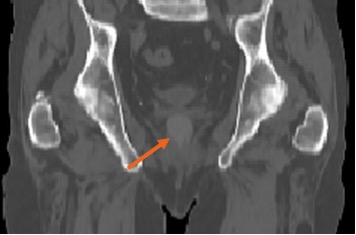}
\caption{}
\label{fig:hydro}
\end{subfigure}
\caption{Examples of OOD Scenarios in radiotherapy contouring: (a) Femur implants cause image artifacts. (b) Brachytherapy applicator devices distort the anatomy. (c) Hydrogel rectal spacer can significantly alter anatomical contours.}
\label{fig:use_cases}
\end{figure*}

In this application study, we evaluate the integration of uncertainty estimation and OOD detection within the framework of a clinical solution from Varian, a Siemens Healthineers company, designed for OAR contouring in radiotherapy. The study leverages a CT dataset comprising scans of the pelvic region, each resampled to an isometric resolution of \SI{1}{\milli\meter} per voxel. Segmentation annotations for six organs -- Bladder, Prostate, Rectum, Femoral Head Left, Femoral Head Right and Seminal Vesicles -- were created in-house by a team consisting of radiologists,  radiation oncologists and trained personal. The primary dataset comprises \num{679} training and \num{20} \emph{control} cases. 
In addition we collect three distinct OOD datasets:

\paragraph{1. Femur Implants Dataset:}  For this study, we created a dataset of \num{13} cases with femoral head implants, extracted from the original OAR contouring training set to simulate an engineering oversight scenario. As depicted in Figure \ref{fig:femur}, femoral head implants create artefacts that impact the anatomical appearance. This deliberate exclusion aims to evaluate the model's capability to identify OOD cases, where critical but potential rare data or patterns might not be part of the training set due to a selection bias.

\paragraph{2. Brachytherapy Dataset:} This dataset encompasses 12 brachytherapy cases, where applicator devices induce anatomical distortions, especially in the rectum as illustrated in Figure \ref{fig:brachy}. The clinical solution is currently not designed or certified for the use in Brachytherapy. The dataset represents a critical OOD scenario where a potential users are unaware of the OAR contouring limitations and unknowingly utilize it in an unsuitable context. 

\paragraph{3. Hydrogel Rectal Spacer Dataset:} We have collected a single case featuring a hydrogel rectal spacer, which significantly impacts anatomical contours, especially within the bladder and rectal regions, as illustrated in Figure \ref{fig:hydro}. Similar to the brachytherapy cases, our clinical solution is not designed for scenarios involving hydrogel rectal spacers, marking it as another OOD case. There is a potential for practitioner oversight that could result in the model being applied in inappropriate contexts. Given its unique nature, this dataset is primarily utilized for qualitative evaluation. \\

These datasets, embodying various OOD scenarios, 
emphasize the critical role of uncertainty estimation in reinforcing system dependability and ensuring the safety of clinical practices. 

\section{Our Method for Epistemic Uncertainty Estimation and OOD Detection}

This section provides details of the architectural design, training protocol, and statistical methods for uncertainty quantification essential to integrate epistemic uncertainty estimation into the radiotherapy contouring workflow. 

\begin{figure}[tbp]
    \centering
    \includegraphics[width=\textwidth]{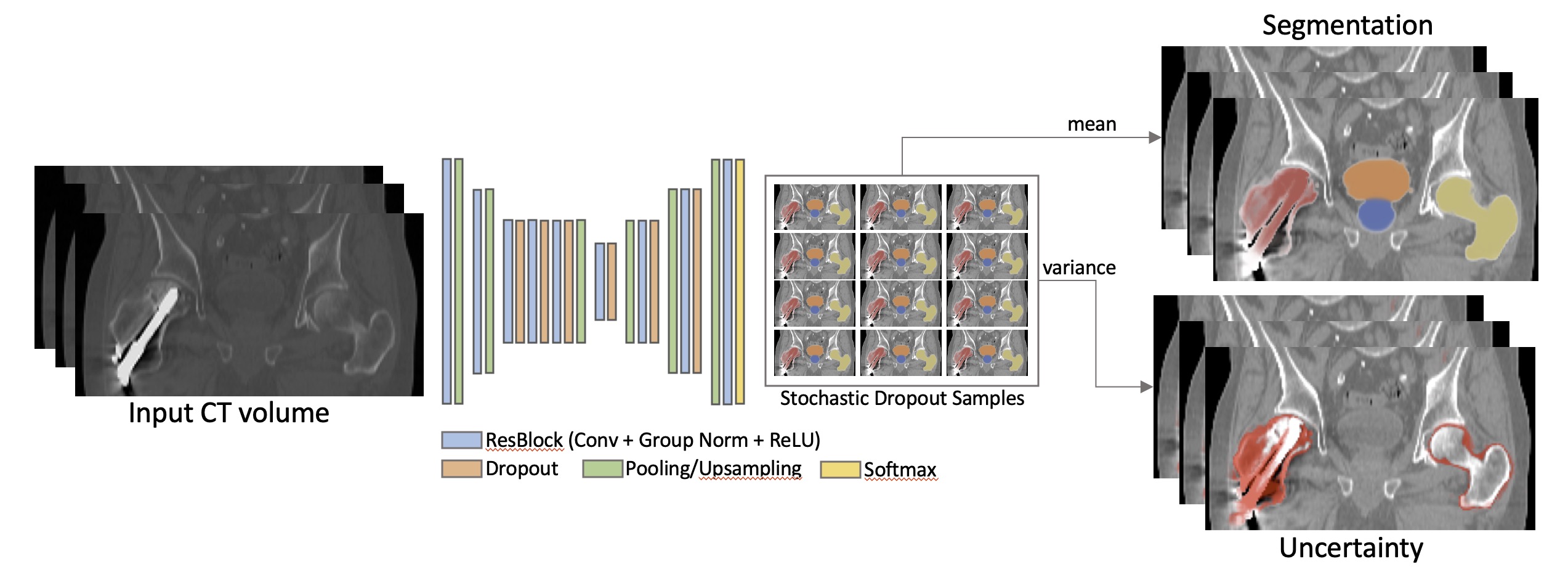}
    \caption{Visualization of our uncertainty model for OAR contouring.}
    \label{fig:model}
\end{figure}

\subsection{OAR Contouring: Architectural Design and Training Protocol}

We approach OAR contouring as a 3D CT image segmentation task, processing 3D volumes (\(H \times W \times T\)) to produce segmentation maps with identical spatial dimensions. 
Our architecture, inspired by ConvNeXt \cite{liu2022convnet} and illustrated in \Cref{fig:model}, adopts a U-Net \cite{ronneberger2015u} style CNN with stages of ResBlocks -- each comprising a 3D convolution, group normalization \cite{wu2018group}, and ReLU activation. Stages include downsampling and upsampling layers, with ResBlock counts of 1,1,3,1 and 1,1,1, respectively, and channel widths doubling after each downsampling and halving upon upsampling, starting from \num{16} channels. For training we minimize a softmax cross-entropy loss utilizing the AdamW optimizer \cite{kingma2014adam,loshchilov2017decoupled}, a batch size of \(\num{8}\), and a \(\num{1e-3}\) initial learning rate, with a polynomial learning rate schedule \cite{teichmann2018convolutional}, weight decay, and mixed-precision training. Augmentation techniques include random rotations (\(\pm \SI{5}{\degree}\)), rescaling (\(\pm \SI{20}{\percent}\)) and gamma correction adjustments. 

\subsection{Methodology for Estimating Epistemic Uncertainty}

We integrate a deep ensemble model \cite{lakshminarayanan2017simple} with Monte Carlo (MC) dropout \cite{gal2015bayesian}, capitalizing on the strengths of both to augment uncertainty estimation within deep convolutional networks. This increases the sample size for uncertainty estimation, which is statistically advantageous offering a more robust measure. The selection of these methods is motivated by several key factors: 1) They specifically model epistemic uncertainty, aligning with our objective of performing OOD detection \cite{kendall2017uncertainties}; 2) Their theoretical foundations are solid and have been well-received within the academic community, ensuring a rigorous approach to uncertainty estimation; 3) Their effectiveness in practical applications has been demonstrated repeatedly, validating their utility in real-world scenarios.

\paragraph{Deep Ensemble Configuration}
\label{par:deep_ens}

Our ensemble model consists of eight base learners, each trained on a subset of the training dataset with $N$ samples to ensure diverse learning perspectives within the ensemble. To optimize the ensemble's coverage and diversity, the dataset is partitioned such that each base learner is exposed to at most $\frac{N}{2} + 1$ of cases, with every case being included in exactly four of the eight base models' training sets. We distribute the samples such that the overlap in training samples between any pair of base learners is limited to at most $\frac{N}{4} + 1$, thereby fostering diversity and maximizing the potential for disagreement among the learners within the specified constraints.

\paragraph{MC Dropout Implementation}

Our MC dropout implementation is inspired by the Bayesian SegNet framework \cite{kendall2015bayesian}. Following the analysis of Kendall et al. \cite{kendall2015bayesian}, we only use dropout during the central stages of our architecture. However, our implementation diverges from Kendall's by inserting MC dropout layers after every ResBlock and not just at the end of each stage, thus more closely aligning with the foundational principles of MC dropout \cite{lakshminarayanan2017simple}. For all Dropouts we utilize 3D dropouts that remove an entire channel in the feature map, addressing the issue that adjacent voxels within feature maps are strongly correlated \cite{tompson2015efficient}. We determine the dropout rate using a hyperparameter search.


\paragraph{Inference and Uncertainty Quantification} For inference, we generate 32 predictions per case by conducting four forward passes for each base learner with MC dropout enabled. An uncertainty heatmap is created by calculating the variance across these predictions for each voxel, yielding a segmentation map with \num{7} channels: one for each organ plus the background. Our post-processing utilizes morphological operations to filter out uncertainty responses at organ boundaries, attributed to annotation variance. An uncertainty score for each organ and case is derived by summing the remaining uncertainty responses within the respective organ channel of the uncertainty map. Uncertainty responses for the background are excluded from further analysis.

\subsection{Statistical Method for OOD Detection}
\label{sec:par_stat_methods}




For statistical threshold estimation and OOD detection we utilize the Mahalanobis distance \cite{lee2018simple}, traditionally applied to network feature distributions for OOD detection, which often necessitates network architecture modifications, such as flattening the final encoder layer \cite{GONZALEZ2022102596} or averaging feature maps \cite{Calli9947059}. These adaptations can lead to potential feature collapse \cite{Lambert10.1007/978-3-031-44336-7_11}. Contrary to these approaches, we directly apply the Mahalanobis distance to uncertainty scores derived from class-conditional distributions, offering a novel method that circumvents the need for architectural changes and avoids the risk of feature collapse.

As outlined in \Cref{par:deep_ens}:\emph{Deep Ensemble Configuration}, our training set is partitioned so each sample is used by only half of the base learners. This partitioning allows for the computation of conservative uncertainty scores $u_i$ using the learners for which the sample was not in the training set. We approximate a class-conditional uncertainty distribution for the in-distribution (ID) training population by estimating $M$ class-conditional Gaussian distributions $\mathcal{N}(\mu_m,\Sigma), m\in[1,M]$ over uncertainty scores $u_i$. Here, $\mu_m=\frac{1}{N}\sum_{i=m}{u_i}$ represents the class-wise means, and $\Sigma=\frac{1}{N}\sum_{m=1}^{M}\sum_{i=m}{(u_i-\mu_m)(u_i-\mu_m)^T}$ is the covariance matrix, capturing shared uncertainties across classes. For each test case with class-wise uncertainty scores $z_i$ the Mahalanobis distance is now given as $\mathcal{D}\mathcal{M}(z_i,\mathcal{N}(\mu{m},\Sigma)) = \sqrt{\sum_{m=1}^{M}(z_i-\mu_{m})^T\Sigma^{-1}(z_i-\mu_{m})}$. An additional advantage is that the Mahalanobis distance follows a $\chi^2$ distribution \cite{https://doi.org/10.1002/cem.2680}, with degrees of freedom equal to the number of features (foreground classes in this case), allowing for the computation of a critical value for OOD detection at a specific significance level without necessitating OOD or ID test samples.

\section{Experimental Evaluation}

In this section, we assess the effectiveness of our Bayesian epistemic uncertainty model in discriminating between ID and OOD cases in the OAR contouring workflow through qualitative analysis and quantitative assessment using the statistical methods discussed in \Cref{sec:par_stat_methods}.

\begin{figure*}[t]
\centering
\begin{subfigure}[t]{\textwidth}
\centering
\includegraphics[width=\textwidth]{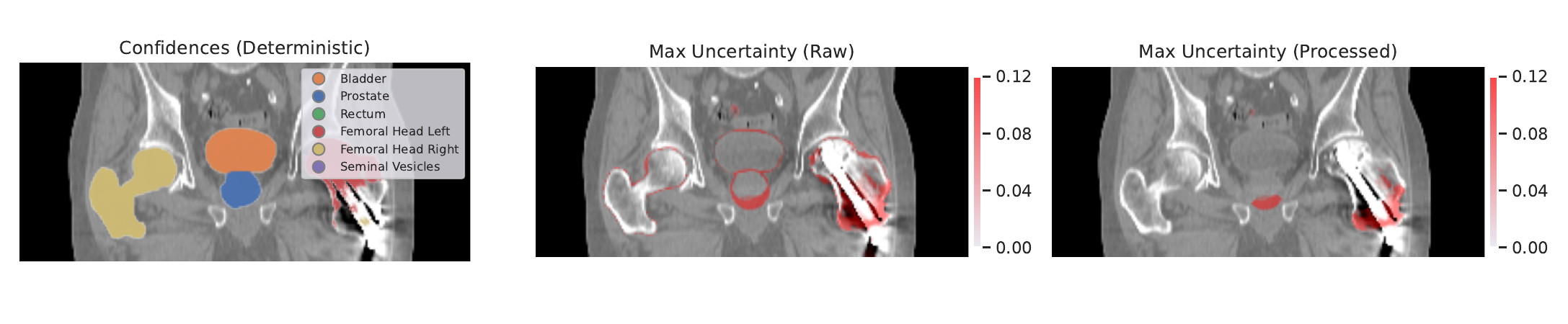}
\caption{Femur Implant: High uncertainty response at the femur implant.}
\label{fig:femurEx}
\end{subfigure}%
~

\begin{subfigure}[t]{\textwidth}
\centering
\includegraphics[width=\textwidth]{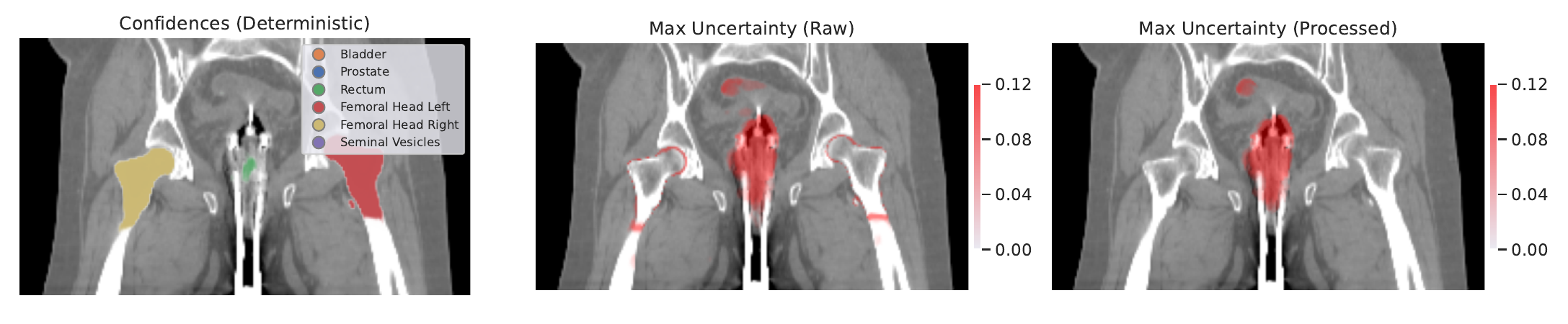}
\caption{Brachytherapy: High uncertainty at rectum and prostate.}
\label{fig:brachyEx}
\end{subfigure}%
~

\begin{subfigure}[t]{\textwidth}
\centering
\includegraphics[width=\textwidth]{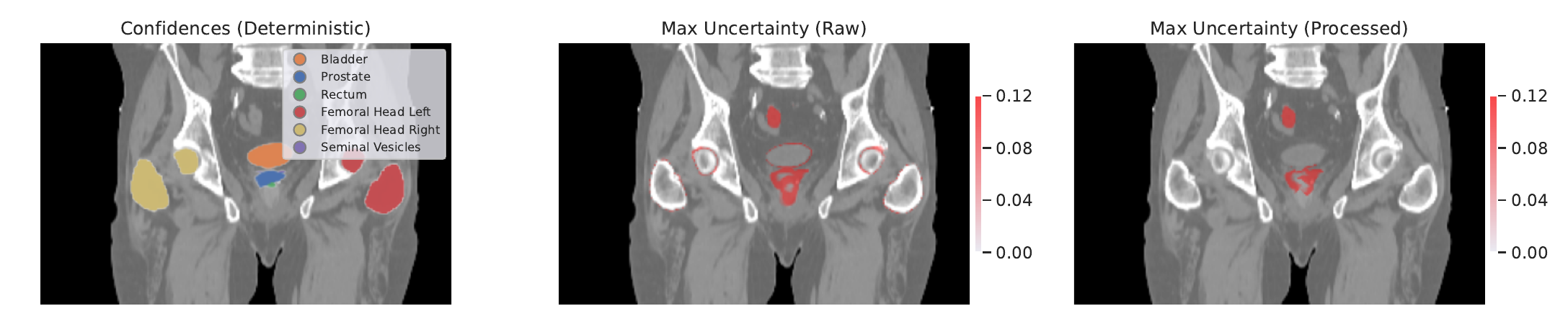}
\caption{Hydrogel Spacer: High uncertainty at the rectum.}
\label{fig:hydroEx}
\end{subfigure}
\caption{Visualization of results for OOD datasets with prediction (left), raw maximum uncertainties (center), processed maximum uncertainties (right). Maximum uncertainty is computed across all foreground classes (organs).}
\label{fig:outputEx}
\end{figure*}

\begin{figure*}[t]
\centering
\begin{subfigure}[t]{0.33\textwidth} 
\centering
\includegraphics[width=\textwidth]{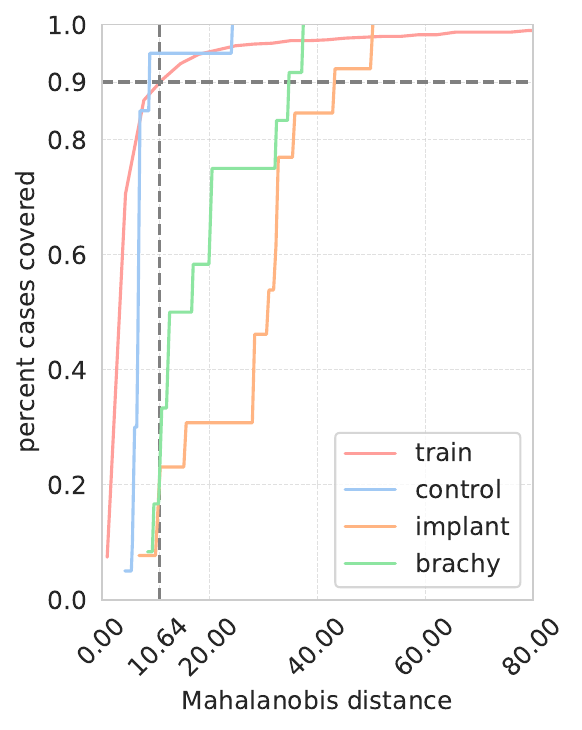}
\caption{}
\label{fig:md_cumulative_curves}
\end{subfigure}%
~
\begin{subfigure}[t]{0.33\textwidth}
\centering
\includegraphics[width=\textwidth]{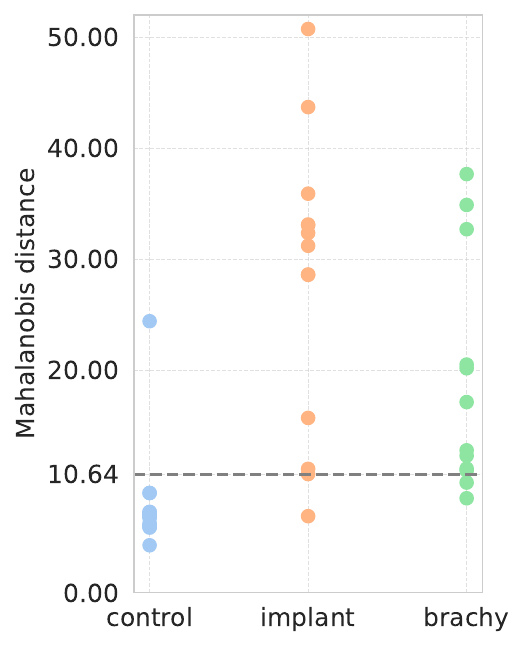}
\caption{}
\label{fig:md_case_values}
\end{subfigure}%
~
\begin{subfigure}[t]{0.33\textwidth}
\centering
\includegraphics[width=\textwidth]{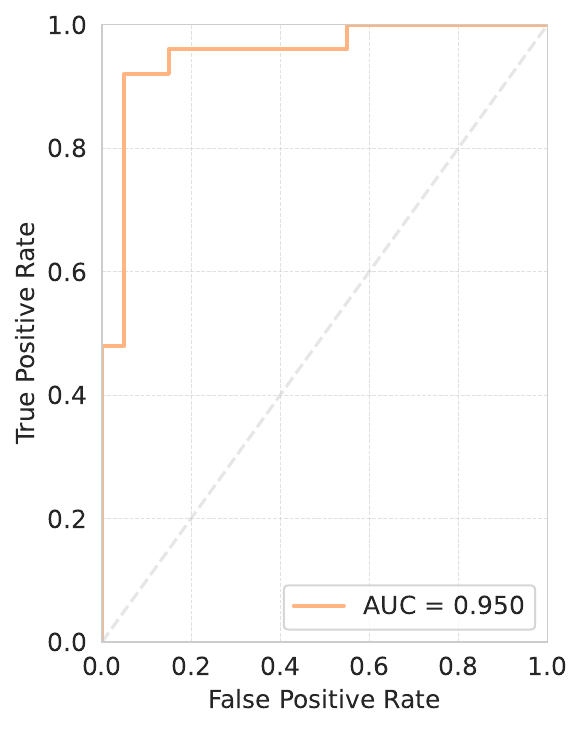}
\caption{}
\label{fig:md_roc_curve}
\end{subfigure}
\caption{Statistical analysis: (a) Cumulative class-conditional Mahalanobis distance curves, x-axis is truncated. (b) Scatterplot of Mahalanobis distance scores for each dataset including OOD threshold. (c) ROC curve for OOD detection.}
\label{fig:md_analysis}
\end{figure*}

\subsection{Qualitative Analysis}

\Cref{fig:outputEx} showcases a representative case from each OOD dataset, illustrating confidence predictions, model uncertainty responses, and post-processed uncertainty for each scenario. Confidence levels are derived from the original deterministic contouring model, not the mean of the Bayesian model, and are visualized as soft-confidences using the alpha channel of the image overlay. A fully opaque pixel in the color of the corresponding class indicates a strong positive prediction by the deterministic model. Conversely, visibility of the background image signifies low confidence in any class, implying a high certainty of background rather than organ presence.


\paragraph{Femur Implants} Figure \ref{fig:femurEx} shows a case with an implant in the left femur head, where an analysis of confidence levels reveals high-confidence false positives and false negatives in the lower region of the left femur head. No low-confidence predictions are observed in this area, indicating that the model generates high-confidence predictions for this OOD case. Notably, our Bayesian model detects a significant uncertainty response in the lower part of the left femur head, accurately identifying the area where predictions are unreliable.

\paragraph{Brachytherapy Cases} Figure \ref{fig:brachyEx} shows a case where a brachytherapy applicator deforms pelvic anatomy, leading to high-confidence false negative predictions for the bladder and prostate. This indicates that the model erroneously predicts with high confidence that these organs are absent. Crucially, our uncertainty model accurately flags areas where model predictions are unreliable due to anatomical distortion caused by the brachytherapy applicator. 


\paragraph{Hydrogel Rectal Spacer} Figure \ref{fig:hydroEx} illustrates a case with a hydrogel spacer inserted between the prostate and rectum, causing deformation of both organs. Despite these alterations, the model produces high-confidence predictions for the prostate, including high-confidence false positives in the spacer's vicinity, mistaking it for prostate tissue. Our approach effectively identifies regions where OAR predictions are affected by the anatomical changes due to the hydrogel spacer, evidencing the model's ability to detect uncertainty in scenarios where organ contours are modified by medical interventions.\\

In summary, our qualitative evaluation shows that model confidences alone do not paint the full picture, especially in OOD scenarios where confidence levels often misrepresent model certainty and fail to provide a reliable indicator of model performance. Conversely, our Bayesian uncertainty approach offers a more accurate depiction of model reliability, effectively identifying areas where predictions are uncertain.

\subsection{Quantitative Analysis}

We use the class-conditional uncertainty distribution based in the training data as discussed in \Cref{sec:par_stat_methods} and estimate the Mahalanobis distance from this distribution for each sample in the ID (training, control) and OOD (implant, brachy) datasets. Using the train distribution we estimate an OOD threshold of \num{10.64} utilizing the $\chi^2$ table \cite{https://doi.org/10.1002/cem.2680} for an application-specific critical value of \num{0.9} and degrees of freedom equal to \num{6} (number of foreground classes). Figure \ref{fig:md_cumulative_curves} illustrates the threshold selection. Note that only the training set is required for the threshold estimation, no OOD or control samples are utilized.

\Cref{fig:md_case_values} illustrates the OOD-detection efficacy of our threshold on the ID and OOD test sets. The threshold accurately classifies \num{95.0}\% of control cases as ID (Specificity), and identifies \num{92.3}\% of Implant cases and \num{83.3}\% of Brachy cases as OOD (Sensitivity). The choice of a critical value, and thus the Sensitivity-Specificity trade-off, may vary based on specific application needs. For a comprehensive evaluation, \Cref{fig:md_roc_curve} showcases our method's threshold-independent performance through an ROC curve, derived from class-conditional Mahalanobis distances across the test dataset. With ID (control) class = $0$ and OOD (brachy, implant) class = $1$, we achieve an AUC-ROC of $0.95$. This demonstrates that our distance-based metric, derived from the underlying multivariate uncertainty distribution, effectively discriminates OOD samples with statistical significance.


\section{Conclusion \& Future Work}

In this study, we have explored the integration of epistemic uncertainty estimation within the radiotherapy contouring workflow, particularly focusing on enhancing the detection of OOD scenarios. Our investigation was motivated by the critical need to improve the safety and reliability of OAR models in radiotherapy, where the contouring precision for target structures and OAR is paramount for effective treatment planning and the safety of the patient. We were able to demonstrate the effectiveness and utility of epistemic uncertainty estimation in identifying instances where model predictions are unreliable. Our approach achieved an AUC-ROC of 0.95 for OOD detection, with a specificity of \num{0.95} and a sensitivity of \num{0.92} for implant cases. These findings highlight the potential of uncertainty estimation as an early warning system for cases requiring additional attention during requiring expert review. By doing so we also address the previously identified gaps in the current research landscape \cite{zou2023review}, namely the lack of ground truth data and tasks as well as the need for empirical evaluations.

Looking forward, our study lays the groundwork for further research into the integration of uncertainty analysis into the radiotherapy workflow. The promising results from our initial feasibility study suggest that epistemic uncertainty estimation can significantly enhance the accuracy, safety, and clinical applicability of deep learning models in radiotherapy planning. Future work will focus on expanding the range of OOD scenarios explored, refining our methodology, and conducting empirical comparisons. Additionally, we encourage the academic community to develop benchmarks and similar tasks based on public data, which can be used for quantitative comparisons and empirical evaluations of various methods. There remains a significant gap in the comparative analysis of uncertainty estimation methods, as highlighted in \cite{zou2023review}. Our work provides meaningful, clinically relevant tasks that the community can use to quantitatively evaluate methods. We hope that our work and results highlight the importance of further research in the space of uncertainty estimation for medical imaging, driving advancements in this critical area.

\begin{credits}

\subsubsection{Disclaimer} The information in this paper is based on research results that are not commercially available. Future commercial availability cannot be guaranteed.

\subsubsection{\discintname}
The authors are employed by Siemens Healthineers, and the research was fully funded by Siemens Healthineers. The authors have no further competing interests to declare that are relevant to the content of this article.

\end{credits}

\bibliographystyle{splncs04}
\bibliography{miccai}

\end{document}